\documentstyle[aps,prb,epsf,floats]{revtex}

\begin{document}


 \twocolumn[{
 \widetext

\title{
A density matrix renormalisation group algorithm for quantum lattice systems with a large number of states per site}

\author{
R. J. Bursill$^*$
}

\address{
School of Physics, University of New South Wales, Sydney, NSW 2052,
Australia
}

\maketitle

 \mediumtext
 { \centering

\begin{abstract}

A variant of White's density matrix renormalisation group scheme which is
designed to compute low-lying energies of one-dimensional quantum lattice
models with a large number of degrees of freedom per site is described.
The method is tested on two exactly solvable models---the spin-1/2
antiferromagnetic Heisenberg chain and a dimerised $XY$ spin chain.
To illustrate the potential of the method, it is applied to a model of
spins interacting with quantum phonons. It is shown that the method
accurately resolves a number of energy gaps on periodic rings which
are sufficiently large to afford an accurate investigation of critical
properties via the use of finite-size scaling theory.

\end{abstract}

 }

\pacs{PACS numbers: }

 }] \narrowtext

\section{Introduction}

\label{introduction}

Since its inception, White's density matrix renormalisation group (DMRG) method \cite{white} has proven to be a powerful, robust and portable numerical method for calculating properties such as excitation energies and correlation functions of low-dimensional quantum lattice models in condensed matter physics \cite{review}. The first applications of the method were to the calculation of static, zero temperature properties of one-dimensional models with short range interactions \cite{spin1,kondo,fermion_ladders,spin_ladders,bursill1}, but the method has been extended to include disordered systems \cite{disorder}, dynamical properties \cite{dynamics}, classical two-dimensional systems \cite{nishino}, finite temperature properties \cite{bursill2}, non-hermitian systems \cite{marston}, and systems with long range interactions \cite{long_range}. Furthermore, pioneering extensions to two spatial dimensions \cite{2D_fermions,CAVO,msdmrg,white_2D,jongh} have rendered the method competitive with any other method for key
two-dimensional models such as the Hubbard model \cite{msdmrg,compare}. The domain of applicability of the method continues to grow with recent applications to wetting phenomena \cite{wetting} and quantum chemistry \cite{quantum_chemistry}.

In principle, little modification is needed in order to apply the DMRG to models with a large number of degrees of freedom per lattice site, such as models with bosonic degrees of freedom, an example being electron-phonon models. As with exact diagonalisation calculations on small clusters, it is possible to truncate the single-site Hilbert space (e.g., by limiting the phonon number in an
electron-phonon model) to the extent where the error in ignoring the shed states is negligible. In practice this will generally mean that the ``added sites'' in the DMRG calculation \cite{white} will contain considerably more states than for a spin or fermion model, but calculations have been performed in this manner for the Bose-Hubbard model \cite{bose_Hubbard}, a model involving spins interacting with phonons \cite{caronxy}, as well as pure acoustic phonons \cite{acoustic_phonons}. For some models, however, the number of states required to accurately represent a single site can begin to become comparable to the number used in representing a whole block (i.e.\ the ``system'' or ``environment'' block \cite{white}), and this straightforward implementation of the DMRG can become inefficient or unworkable.

White and co-workers have developed two methods for dealing with the single-site Hilbert space. In one approach, $2^n$ truncated phonon Hilbert space levels are represented by $n$ spinless fermions and a standard ``sweeping'' method \cite{white} is employed to add just one fermion site at a time rather than a whole phonon degree of freedom. This method was used to study the polaron problem (a single electron interacting with a lattice of vibrating atoms) in the one- and
two-dimensional Holstein models \cite{jeckelmann}. A more promising approach, that of ``local Hilbert space reduction'' \cite{zhang}, involves finding a highly efficient single-site basis which can be used as an alternative to simple Hilbert space truncation when performing DMRG or exact diagonalisation studies. This is achieved by taking a small system that can readily be exactly diagonalised, e.g.\ a system with 4 sites, and then using the projection operator for the ground state and/or some low-lying excitations to define a reduced density matrix for a single site, by tracing the projection operator over the degrees of freedom of all but one of the sites \cite{zhang}. The resulting basis was shown to be very efficient in exact diagonalisation studies of four- and six-site half-filled Hubbard-Holstein systems, in that the number of states required to accurately represent the site was very small \cite{zhang}.

In this paper we present a DMRG scheme for one-dimensional quantum lattice systems with a large number of degrees of freedom per site which has some similarities to the local Hilbert space reduction scheme \cite{zhang}, and also to Wilson's computer renormalisation group method \cite{wilson}. We will call the method the ``four-block method'', as it uses four blocks rather than the two blocks used in the standard DMRG. The four-block method has been used to perform a study of an electron-phonon model---the spinless fermion Holstein model \cite{holstein}. In this study the phase boundary separating the metallic and insulating phases of the model was determined with high accuracy, and very good agreement was obtained with analytical results which become exact in the nontrivial strong coupling limit \cite{holstein}. The four-block method is described in Section \ref{method}. The accuracy is then tested on solvable models. Finally, the potential of the method to study systems with a large number of degrees of freedom per site is demonstrated by presenting some convergence results for the $XY$ spin-Peierls model.

\section{The four-block method}
\label{method}

The four-block method is illustrated schematically in Fig.\ \ref{schematic}. A calculation commences with a ring of four sites. A basis for the ring is the product of 4 copies of a single-site basis i.e.\
\begin{equation}
B_{\text{ring}} \equiv
B_{\text{s}} \otimes B_{\text{s}} \otimes B_{\text{s}} \otimes B_{\text{s}},
\end{equation}
where
$
B_{\text{s}} = \{ | n \rangle: n = 1, 2, 3, \ldots \}
$
is a single-site basis. For example, for a pure phonon system we might choose $| n \rangle$ to be the $n$th phonon level. The ring basis is reduced down to a finite set by limiting the size of the single-site basis. That is, $ B_{\text{ring}} $ is replaced by
\begin{eqnarray}
B_{\text{ring}}^{(m)}
& \equiv &
B_{\text{s}}^{(m)} \otimes B_{\text{s}}^{(m)} \otimes B_{\text{s}}^{(m)}
\otimes B_{\text{s}}^{(m)}
\\
& = &
\{ | n_1 \rangle \otimes | n_2 \rangle \otimes | n_3 \rangle \otimes
| n_4 \rangle:
\nonumber
\\
& &
n_1, n_2, n_3, n_4 = 1, \ldots, m  \}
\end{eqnarray}
where 
$
B_{\text{s}}^{(m)} \equiv \{ | n \rangle: n = 1, \ldots, m \}
$
is a truncated single-site Hilbert space. The cutoff $m$ is chosen so that, for the purpose of finding low-lying excitations of the ring Hamiltonian, the error in ignoring the shed states is negligible. The ring Hamiltonian is diagonalised by a sparse matrix method to produce the ground state
$ | \psi \rangle $.

The next step is to consider a ``system'' block $A$ consisting of sites 1 and 2, as shown in Fig.\ \ref{schematic}. A truncated basis for this block is
\begin{eqnarray}
B_A & \equiv &
B_{\text{ring}}^{(m)} \otimes B_{\text{ring}}^{(m)}
\\
& = &
\{ | n_1 \rangle \otimes | n_2 \rangle: n_1, n_2 = 1,\ldots, m \}.
\end{eqnarray}
A reduced density matrix $\rho_A$ is defined for block $A$ by integrating out the ``environmental'' degrees of freedom, $n_3$ and $n_4$, from the ground state projection operator
$ | \psi \rangle \langle \psi | $, viz.\
\begin{eqnarray}
& &
\langle n_1' | \otimes \langle n_2' | \rho_A | n_1 \rangle
\otimes | n_2 \rangle
\equiv
\nonumber
\\
& &
\sum_{n_3, n_4 = 1}^{m}
\psi^*_{n_1' n_2' n_3 n_4} \psi_{n_1 n_2 n_3 n_4},
\end{eqnarray}
where
$ \psi_{n_1 n_2 n_3 n_4} \equiv \left\{ \langle n_1 | \otimes
\langle n_2 | \otimes \langle n_3 | \otimes \langle n_4 | \right\}
| \psi \rangle $. Next, $\rho_A$ is diagonalised yielding eigenvalues $\omega_n$ and eigenvectors $|n\rangle\rangle$. The density matrix eigenvalues $\omega_n$ are real and positive and sum to unity, as $\rho_A$ is a probability matrix, viz.\
\begin{eqnarray}
1 > \omega_1 \geq \omega_2 \geq \ldots \geq \omega_{m^2} & \geq & 0;
\\
\sum_{n = 1}^{m^2} \omega_n = 1.
\end{eqnarray}
At this stage a cutoff $\tilde{m} \leq m^2$ is chosen and a new truncated basis $\tilde{B}_A^{(\tilde{m})}$ is developed for the block $A$ from the $\tilde{m}$ ``most important'' density matrix eigenstates, i.e.\
\begin{equation}
\tilde{B}_A^{(\tilde{m})} \equiv \{ |n\rangle\rangle: 
n = 1,\ldots,\tilde{m} \}
\end{equation}

The procedure of forming a ring, or ``superblock'', is then repeated, this time using four two-site blocks (copies of the system block) instead of four single sites, as shown in Fig.\ \ref{schematic}. $\tilde{m}$, $\tilde{B}_A^{(\tilde{m})}$ and $|n\rangle\rangle$ play the r\^{o}les of $m$, $B_{\text{site}}^{(m)}$ and $|n\rangle$, and again the superblock Hamiltonian is diagonalised for the ground state $|\psi\rangle$; a new system block, consisting of two blocks (or 4 sites) is generated, and a reduced density matrix is defined. The procedure is iterated, with the lattice (or superblock) size doubling at each iteration, as is the case with Wilson's computer renormalisation group method \cite{wilson}. The method also has some similarity with the local Hilbert space reduction technique \cite{zhang} in that initially a local Hilbert space is defined for a two-site block by means of a reduced density matrix.

At first glance the four-block method might appear difficult to implement as the size of the superblock Hilbert space grows (with the number of states retained per block, $m$) as $m^4$ rather than $m^2$ for the standard DMRG method \cite{white}. Fortunately, there are a number of steps that can be taken to reduce the computer resource requirements.

\begin{enumerate}

\item
In defining a truncated basis for the superblock, rather than choosing $m$ and taking
\begin{eqnarray}
B_{\text{super}}^{(m)} = \left\{
|n_1\rangle\rangle \otimes |n_2\rangle\rangle \otimes 
|n_3\rangle\rangle \otimes |n_4\rangle\rangle:
\right.
\nonumber
\\
\left.
n_1, n_2, n_3, n_4 = 1,\ldots,m
\right\},
\end{eqnarray}
we choose a cutoff $ 0 < \epsilon \leq 1 $ and define the truncated superblock basis according to
\begin{eqnarray}
B_{\text{super}}^{(\epsilon)} = \left\{
|n_1\rangle\rangle \otimes |n_2\rangle\rangle \otimes 
|n_3\rangle\rangle \otimes |n_4\rangle\rangle:
\right.
\nonumber
\\
\left.
\omega_{n_1} \omega_{n_2} \omega_{n_3} \omega_{n_4} \geq \epsilon
\right\}.
\end{eqnarray}
That is, the superblock basis is taken to be the set of all fourfold products of density matrix eigenstates such that the product of the corresponding density matrix eigenvalues is $\epsilon$ or greater. Thus, {\em $\epsilon$ is the single truncation parameter which determines the accuracy of the calculation}. Denoting the size of the superblock Hilbert space by
$ M^{(\epsilon)} \equiv \left| B_{\text{super}}^{(\epsilon)} \right| $,
we note that $ M^{(\epsilon)} $, or the accuracy of the calculation, increases as $ \epsilon $ is decreased. Typical values of $ \epsilon $ range from $10^{-10}$ to $10^{-25}$, and allow the study of systems with very large effective values of $m$, i.e.\ some hundreds of states can be retained per block. For a given accuracy requirement, the advantage (in terms of CPU and memory) of using $ B_{\text{super}}^{(\epsilon)} $
instead of $ B_{\text{super}}^{(m)} $ is some orders of magnitude. This is because many of the states in $ B_{\text{super}}^{(m)} $ contain two or more block states with low density matrix eigenvalues and thus have very low probability. These states are not considered when $ B_{\text{super}}^{(\epsilon)} $ is used. This approach to Hilbert space truncation can also be applied in the standard DMRG algorithm. A modification along these lines is given in \cite{lepetit}.

\item
Because the four blocks that make up the superblock are identical, use can be made of the translational and/or reflection symmetries in reducing the CPU time required to act the superblock Hamiltonian
${\cal H}_{\text{super}}$ on a state. This operation, needed for the sparse diagonalisation of ${\cal H}_{\text{super}}$, is the most CPU intensive operation in the algorithm.

\item
The algorithm has a natural vectorisation. Because any given term in ${\cal H}_{\text{super}}$ only connects two of the four blocks, inner loops can be taken over the states of the inactive blocks. For example, consider a term in ${\cal H}_{\text{super}}$, ${\cal H}_{12}$, which connects blocks 1 and 2. Taking an initial superblock state
$ |n_1\rangle \otimes |n_2\rangle \otimes |n_3\rangle \otimes |n_4\rangle $,
the most general final state (under the action of
${\cal H}_{12}$) is 
$ |n_1'\rangle \otimes |n_2'\rangle \otimes |n_3\rangle \otimes |n_4\rangle $, i.e.\ the indices for blocks 3 and 4 are unchanged. Making outer loops over $n_1$, $n_2$, $n_1'$ and $n_2'$, the matrix element
$
\chi \equiv \left\langle n_1' \left| \otimes \left\langle n_2' \left|
{\cal H}_{12} \right| n_1 \right\rangle \otimes \right| n_2 \right\rangle
$
is calculated or read in from storage. An inner loop in which $\chi$ is repeatedly reused can be performed over $n_3$ and $n_4$ (melded into a single index). The superblock states can be ordered in such a way that all memory access (to arrays representing superblock states) in the inner loops is contiguous.

\end{enumerate}

\section{Accuracy tests for the Heisenberg chain}

The four-block method has been tested on the $S = 1/2$ antiferromagnetic Heisenberg spin chain
\begin{equation}
{\cal H} = 2 \sum_{i=1}^{N} S_i . S_{i+1},
\end{equation}
where $S_i$ is the spin-1/2 operator for site $i$ and a periodic ring of $N$ sites is assumed. This model is exactly solvable by Bethe ansatz \cite{bethe} and in particular, exact results are available for the ground state energy $E_{\text{GS}}$, and the singlet $\Delta_{\text{ss}}$ and triplet $\Delta_{\text{st}}$ gaps on finite periodic rings \cite{finite_rings}.

The states associated with these gaps are found by making use of the parity (spin-flip) operator
\begin{equation}
\hat{T}: S_i^z \longrightarrow -S_i^z,\;\;
         S_i^+ \longrightarrow S_i^-,
\label{spinflip}
\end{equation}
and utilising a projection operator of the form \cite{white}
$$
\frac{1}{3} \left\{
\left| \psi_0^{(0)} \right\rangle \left\langle \psi_0^{(0)} \right| +
\left| \psi_0^{(1)} \right\rangle \left\langle \psi_0^{(1)} \right| +
\left| \psi_1^{(0)} \right\rangle \left\langle \psi_1^{(0)} \right|
\right\},
$$
instead of the ground state projection operator to define the density matrix, where $ \left| \psi_0^{(0)} \right\rangle $,
$ \left| \psi_0^{(1)} \right\rangle $ and
$ \left| \psi_1^{(0)} \right\rangle $ denote the ground  state and the first excited triplet and singlet states respectively.

Results for $E_{\text{GS}}$, $\Delta_{\text{ss}}$ and $\Delta_{\text{st}}$ for various values of $\epsilon$ and $N = 32$, 64 and 128 are given in Table \ref{Heisenberg}. The high accuracy for
$N = 32$ is to be expected as a substantial fraction of the complete Hilbert space is retained in this case. For $N = 128$ the singlet and triplet gaps are resolved to within around 0.1\%. Such accuracy for periodic rings in critical models where the gaps vanish in the thermodynamic limit makes the
four-block method potentially useful for finite-size scaling studies. The four-block method, like the standard DMRG, is variational in that total energies improve monotonically with decreasing $\epsilon$ (or increasing Hilbert space size $M(\epsilon)$). Use can be made of $1 / M(\epsilon)$ to extrapolate to the $\epsilon \rightarrow 0$ limit. Results of such linear, two-point extrapolations for the $N = 64$ and 128 cases are included in Table \ref{Heisenberg} and generally improve the results from the largest value of $M(\epsilon)$. Note, however, that the extrapolations are not variational in general, nor are any of the results for the gaps, which are the difference of two total energies.

\section{Use of the translation operator}

A feature of the four-block method is that the reduced Bloch symmetry of the four-block ring can be used to explicitly target states in momentum sectors other than $k = 0$ and $k = \pi$. For example, by constructing superblock states with a phase of $-1$ under shifts:
$$
|n_1\rangle \otimes |n_2\rangle \otimes |n_3\rangle \otimes
|n_4\rangle \otimes \rightarrow |n_2\rangle \otimes |n_3\rangle \otimes |n_4\rangle \otimes |n_1\rangle,
$$ 
the $k = 2 \pi / N$ symmetry sector can be targeted directly.

For example, we consider the exactly solvable, dimerised $XY$ spin chain:
\begin{equation}
{\cal H} = 2 \sum_{i = 1}^N
\left( S_i^x S_{i+1}^x + S_i^y S_{i+1}^y \right)
\left( 1 + (-1)^i \delta \right),
\end{equation}
where $\delta$ is the dimerisation parameter.
The exact excitation spectrum in any momentum sector is readily obtained for finite lattices \cite{bulaevskii}. Table \ref{dimerisedXY} shows the convergence of the four-block method for the energy gap
$\Delta_{\pi / 16}$, from the ground state to the lowest excitation in the $k = \pi / 16$ sector, for the $N = 64$ site ring with
$\delta = 0.2$. Note that $\Delta_{\pi / 16}$ is resolved to within around 0.01\%. The reduced Bloch symmetry may prove to be useful in mapping out the excitation spectrum for, say, a 64 site lattice.

\section{Application of the four-block method to systems with a large number of degrees of freedom per site}

In the above sections we have shown that the four-block method can provide accurate determinations of energy gaps in spin models on large, periodic rings. However, the real utility in the method lies in its ability to deal with systems with a large number of degrees of freedom per site. To illustrate this we consider the $XY$ spin chain interacting with dispersionless quantum phonons \cite{caronxy}. Here we are mainly concerned with demonstrating the convergence of the four-block method, rather than performing a comprehensive study of the model. The Hamiltonian is given by
\begin{eqnarray}
{\cal H} & = & \sum_i
\left[
1 + g \left( b_{i+1}^\dagger + b_{i+1} - b_i^\dagger - b_i \right)
\right.
\nonumber
\\
& &
\times \;
\left.
\left( S_i^x S_{i+1}^x + S_i^y S_{i+1}^y \right)
\right]
+ \omega \sum_i b_i^\dagger b_i,
\end{eqnarray}
where $b_i$ destroys a phonon of frequency $\omega$ and $g$ is the
spin-phonon coupling. This model has been studied by Caron and Moukouri using the standard DMRG method \cite{caronxy}. The model undergoes a Kosterlitz-Thouless (K-T) transition at some critical coupling
$g_{\text{c}}$ from a Luttinger liquid phase $(g < g_{\text{c}})$ with gapless excitations to a gapped, dimerised phase $(g > g_{\text{c}})$ with a doubly degenerate ground state. In \cite{caronxy} the gap $\Delta$ in the dimerised region is determined as a function of $g$ in the thermodynamic limit by performing DMRG calculations on large lattices. This data is fitted to Baxter's K-T form \cite{baxter}
\begin{equation}
\Delta = \frac{a}{\sqrt{g^2 - g_{\text{c}}^2}}
\exp \left( - \frac{b}{\sqrt{g^2 - g_{\text{c}}^2}} \right),
\label{KTform}
\end{equation}
in order to determine $g_{\text{c}}$.

Here we show that the four-block method can be used to accurately calculate a number of energy gaps on finite, periodic rings. The examination of the crossover of these gaps allows an accurate determination of the critical point for models with a K-T transition \cite{nomura1,nomura2,holstein}. The good quantum numbers for this model that can be exploited by the four-block method are the total $z$ spin, $S^z_{\text{T}}$, and, in the $S^z_{\text{T}} = 0$ sector, the spin-flip symmetry (\ref{spinflip}). In addition, the reduced Bloch symmetry and reflection symmetry can be used. In addition to the ground state energy
$E_{\text{GS}} = E_0(S^z_{\text{T}} = 0,\hat{T} = 1)$, three gaps are considered. In the notation of Nomura, these are: the doublet gap:
$\Delta_{\text{doublet}} \equiv E_0(S^z_{\text{T}} = \pm 1) - E_{\text{GS}}$;
the dimer gap:
$\Delta_{\text{dimer}} \equiv E_1(S^z_{\text{T}} = 0,\hat{T} = 1) - E_{\text{GS}}$;
and the N\'{e}el gap:
$\Delta_{\text{N\'{e}el}} \equiv E_0(S^z_{\text{T}} = 0,\hat{T} = -1) - E_{\text{GS}}$ \cite{footnote}.

The first stage of the four-block method requires the exact diagonalisation of a four-site ring. Convergence results for the various gaps in the four-site system are given in Table \ref{XYSP4} for the $\omega = 10$, $g = 2.4$ case. Note that in this case convergence with $m$, the number of bare phonon modes retained per site, is very rapid. If $\omega$ (the energy needed to create a phonon excitation) is decreased and/or the coupling $g$ is increased, the value of $m$ required for convergence will increase. In calculations performed on various electron-phonon models \cite{holstein,bursillX} the required value has not proved prohibitive i.e.\ $m \leq 30$. If this stage of the calculation does present a problem then global Hilbert space truncation can be used, i.e., rather than placing a limit $m$ on the phonon number for each site, the sum of the phonon numbers from all sites is restricted. This simple step dramatically reduces the size of the Hilbert space required for convergence. Failing this, the local Hilbert space reduction method of White and co-workers \cite{zhang} can be used to build an efficient basis for the f-site ring by starting with a two-site system and forming a single-site density matrix. Even the two-site calculation can be made more efficient by using a coherent state basis rather than simply bare phonon states.

The convergence with $\epsilon$ of subsequent stages of the four-block method---$N = 8$, 16 and 32---is illustrated in Table \ref{XYSP81632}. The convergence is sufficiently rapid that the data can be used in finite-size scaling studies (the gaps in the $N = 32$ case are resolved to within around 0.01\%). As an example we consider the critical coupling $g_{\text{c}}$. $g_{\text{c}}$ can be obtained as the limiting value of $g_{\text{c}}(N)$, where $g_{\text{c}}(N)$ denotes the
finite-size crossover from the gapless spin-fluid phase to the dimerised phase, and is fixed by the condition \cite{nomura2}
\begin{equation}
\Delta_{\text{doublet}} = \Delta_{\text{dimer}}.
\label{crossover_equation}
\end{equation}
That is, the lowest excitation in the fluid phase is the doublet, whose energy gap vanishes in the bulk limit, whilst in the dimer phase the dimer excitation becomes degenerate with the ground state in the bulk limit, whereas the doublet energy gap approaches a non-zero limit, the energy gap $\Delta$ used in \cite{caronxy}.

\begin{figure}[p]
\centerline{\epsfxsize=8.4cm \epsfbox{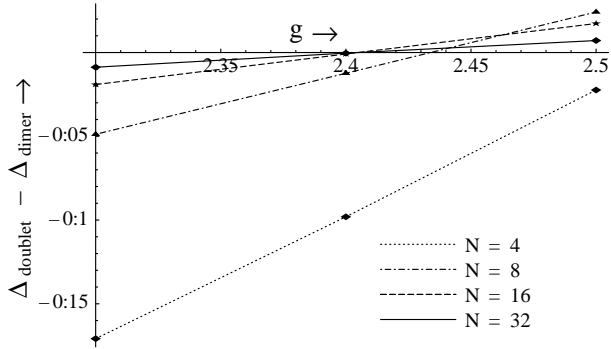}}
\caption{
The difference of the doublet and dimer gaps as a function of $g$ for the $XY$ spin-Peierls model with $\omega = 10$ as determined by the
four-block method for $N = 4$ (dotted line), $N = 8$ (dot-dashed line),
$N = 16$ (dashed line), and $N = 32$ (full line). The crossover point $g_{\text{c}}(N)$ is determined by the condition
$\Delta_{\text{doublet}} = \Delta_{\text{dimer}}$ (the intercepts of the curves with the horizontal axis) and converges rapidly to the critical point $g_{\text{c}}$ as $N \rightarrow \infty$.
}
\label{crossover}
\end{figure}

Plots of $\Delta_{\text{doublet}} - \Delta_{\text{dimer}}$ versus $g$ in the $\omega = 10$ case are given in Fig.\ \ref{crossover} for various values of $N$. A simple quadratic fit of the data in Fig.\ \ref{crossover} gives estimates for $g_{\text{c}}(N)$ which are tabulated in Table \ref{gcofN}. Note the very rapid convergence of
$g_{\text{c}}(N)$ with $N$ \cite{nomura1}. For smaller values of $\omega$ there is strong mixing between fermion-like and phonon-like excitations \cite{holstein} in the dimer and N\'{e}el sectors and it is only for large lattices that $g_{\text{c}}(N)$ converges \cite{holstein}, when the characteristic electron gap, $2 \pi / N$, falls below $\omega$, the energy required to create a phonon excitation.

Taking into account the discretisation and fitting errors, we can safely estimate $g_{\text{c}} = 2.41(3)$. This result is to be compared with the  result $g_{\text{c}} \approx 2.9$, obtained from the phase boundary in \cite{caronxy}. The discrepancy between the two results is probably due to the problematic nature of fitting the infinite system gap $\Delta$ to (\ref{KTform}). That is, three parameters, $a$, $b$ and $g_{\text{c}}$, must be obtained from the non-linear fit, and it is very difficult to determine $\Delta$ accurately near $g = g_{\text{c}}$. This is because $\Delta$ is extremely small for values of $g$ even substantially higher than $g_{\text{c}}$, due to the essential singularity in (\ref{KTform}). Determining such small gaps from
finite-size scaling is very difficult as very large lattices are required in order to observe the crossover from the initial algebraic scaling with $N$ to the exponential scaling expected for gapped systems. This is further complicated by the fact that open, rather than periodic, boundary conditions were used in \cite{caronxy} and by the presence of substantial DMRG truncation error for long chains and small gaps. These factors could well lead to an overestimation of $g_{\text{c}}$.

Finally, we perform a consistency check on our hypothesis that the transition is of the K-T type. Following \cite{nomura1} and \cite{nomura2}, we define parameters $v_0$ and $v_1$ according to
\begin{equation}
E_{\text{GS}} = N \epsilon_{\infty} + \frac{\pi^2 v_0}{6 N} + \ldots,
\end{equation}
and
\begin{equation}
\frac{1}{4} \left[
2 \Delta_{\text{doublet}} + \Delta_{\text{dimer}} +
\Delta_{\text{N\'{e}el}} \right] = \frac{\pi v_1}{N} + \ldots,
\end{equation}
where $\epsilon_\infty$ is the bulk ground state energy per site.
$v_0$ and $v_1$ are extracted from the finite-size scaling of
$E_{\text{GS}}$ and the gaps. Two-point extrapolations of the $N = 16$ and $N = 32$ results give $v_0 / v_1 = 0.9980$ for $g = 2.4 \approx g_{\text{c}}$. This is highly consistent with the result $v_0 / v_1 = 1$ which should hold at the fluid-dimer transition point \cite{nomura1}.

\section{Summary}

In this paper a new variant of White's density matrix renormalisation group (DMRG) method was presented. The algorithm was designed for the performance of finite-size scaling studies of one-dimensional quantum lattice models with a large number of degrees of freedom per site. We call the technique the ``four-block method'', because superblocks consisting of four identical blocks are used as opposed to the standard DMRG method which uses two. The four-block method was shown to recover exact Bethe ansatz results for spin-1/2 Heisenberg rings of up to 128 sites with good accuracy. It was shown that partial use can be made of the Bloch symmetry so that momentum sectors other than $k = 0$ and
$k = \pi$ can be targeted directly.

However, the real utility of the four-block method lies in its ability to treat systems with a large number of degrees of freedom per site such as electron-phonon models. This was demonstrated by applying the method to the $XY$ spin-Peierls model. It was shown that the method accurately resolves a number of finite-system energy gaps in this model, and, using finite-size scaling, the critical coupling was accurately determined for one particular phonon frequency. In future studies the Heisenberg spin-Peierls model will be investigated \cite{bursillX}. Extensions of the four-block method to higher dimensions are also being pursued.

\acknowledgements

I gratefully acknowledge useful discussions with Dr R. McKenzie, Prof.\ C. J. Hamer and Dr T. Xiang. Calculations were performed at The New South Wales Center for Parallel Computing and The Australian National University Supercomputing Facility. I thank Dr R. Standish, Dr D. Singleton and Dr J. Jenkinson for technical support. This work was supported by the Australian Research Council.

\vfill
\eject

\vfill
\eject

\begin{table}[htbp]
\caption{
The ground state energy, $E_{\text{GS}}$, and the singlet and triplet gaps, $\Delta_{\text{ss}}$ and $\Delta_{\text{st}}$, of $N=32$, 64 and 128 site spin-1/2 antiferromagnetic Heisenberg rings calculated using the four-block method for a number of cutoff parameters $\epsilon$. The size of the superblock Hilbert space is $M(\epsilon)$. Exact results are taken from ref.\ \protect\cite{finite_rings}. Two-point extrapolations to the $\epsilon \rightarrow 0$ limit (using $1 / M(\epsilon)$) are included for the $N=64$ and $N=128$ cases.
}
\begin{tabular}{lcllll}
$N$ & $\epsilon$ & $M(\epsilon)$ & $E_{\text{GS}}$ & $\Delta_{\text{ss}}$ & $\Delta_{\text{st}}$ \\
\hline
32  & $10^{-12}$ & 321085  & $-$28.41270950 & 0.42268877 & 0.27638663 \\
32  & $10^{-13}$ & 558920  & $-$28.41293094 & 0.42259417 & 0.27639298 \\
32  & $10^{-14}$ & 931016  & $-$28.41300688 & 0.42256120 & 0.27639284 \\
32  & $10^{-15}$ & 1505096 & $-$28.41303768 & 0.42254578 & 0.27639437 \\
32  & $10^{-16}$ & 2359314 & $-$28.41304884 & 0.42254083 & 0.27639534 \\
32  & $10^{-17}$ & 3605973 & $-$28.41305279 & 0.42253859 & 0.27639562 \\
32  & $10^{-18}$ & 5399963 & $-$28.41305418 & 0.42253776 & 0.27639573 \\
32  & $10^{-19}$ & 8783103 & $-$28.41305472 & 0.42253743 & 0.27639577 \\
32  &    Exact   &   ---   & $-$28.41305488 & 0.42253733 & 0.27639579 \\
64  & $10^{-10}$ & 290849  & $-$56.74279    & 0.20458    & 0.14057    \\
64  & $10^{-11}$ & 616558  & $-$56.74617    & 0.20344    & 0.14051    \\
64  & $10^{-12}$ & 1233916 & $-$56.74755    & 0.20306    & 0.14044    \\
64  & $10^{-13}$ & 2357632 & $-$56.74816    & 0.20287    & 0.14043    \\
64  & $10^{-15}$ & 7567039 & $-$56.74851    & 0.20275    & 0.14041    \\
64  & $\epsilon \rightarrow 0$
                 &   ---   & $-$56.74867    & 0.20270    & 0.14041    \\
64  &    Exact   &   ---   & $-$56.74860    & 0.20271    & 0.14042    \\
128 & $10^{-10}$ & 711104   & $-$113.4476   & 0.10043    & 0.07179    \\
128 & $10^{-11}$ & 1631884  & $-$113.4532   & 0.09948    & 0.07126    \\
128 & $10^{-12}$ & 3482469  & $-$113.4558   & 0.09886    & 0.07107    \\
128 & $10^{-13}$ & 7008910  & $-$113.4570   & 0.09854    & 0.07102    \\
128 & $10^{-14}$ & 13380383 & $-$113.4575   & 0.09839    & 0.07100    \\
128 & $\epsilon \rightarrow 0$
                 &   ---    & $-$113.4581    & 0.09823    & 0.07097   \\
128 &  Exact     &  ---     & $-$113.4585    & 0.09815    & 0.07104   \\
\end{tabular}
\label{Heisenberg}
\end{table}

\begin{table}[htbp]
\caption{
The energy gap $\Delta_{\pi / 16}$ from the ground state to the first excited state in the $k = \pi / 16$ momentum sector for the $N = 64$ site dimerised $XY$ model with dimerisation $\delta = 0.2$ calculated using the four-block method for a number of cutoff parameters $\epsilon$. The size of the superblock Hilbert space is $M(\epsilon)$.
}
\begin{tabular}{lld}
$\epsilon$ & $M(\epsilon)$ & $\Delta_{\pi / 16}$ \\
\hline
$10^{-12}$ & 2055402  &  0.90596 \\
$10^{-13}$ & 3882712  &  0.90476 \\
$10^{-14}$ & 6984756  &  0.90429 \\
$10^{-15}$ & 12074567 &  0.90411 \\
  Exact  &     ---    &  0.90401 \\
\end{tabular}
\label{dimerisedXY}
\end{table}

\begin{table}[htbp]
\caption{
Convergence of various energy gaps at the first stage (exact diagonalisation of a four-site ring) of the four-block method for the $XY$ spin-Peierls model with $\omega = 10$ and $g = 2.40$. $m$ is the number of bare phonon levels retained per site in the truncated basis and
$M(m) = 6 m^4$ is the size of the four-site Hilbert space.
}
\begin{tabular}{lllll}
$m$ & $M(m)$ & $\Delta_{\text{doublet}}$ & $\Delta_{\text{dimer}}$ & $\Delta_{\text{N\'{e}el}}$ \\
\hline
5  & 3750    & 0.767464361243 & 0.865542601399 & 0.941963280212 \\
8  & 24576   & 0.767464370821 & 0.865542584041 & 0.941963270714 \\
12 & 124416  & 0.767464370824 & 0.865542584060 & 0.941963270724 \\
\end{tabular}
\label{XYSP4}
\end{table}

\begin{table}[htbp]
\caption{
Convergence of various energy gaps calculated using the four-block method for the $XY$ spin-Peierls model with $\omega = 10$ and $g = 2.40$ and
$N = 8$, 16 and 32 sites. Here $\epsilon$ is the cutoff parameter which determines the accuracy of the method and $M(\epsilon)$ is the size of the superblock Hilbert space.
}
\begin{tabular}{lcllll}
$N$ & $\epsilon$ & $M(\epsilon)$ & $\Delta_{\text{doublet}}$ & $\Delta_{\text{dimer}}$ & $\Delta_{\text{N\'{e}el}}$ \\
\hline
8  & $10^{-12}$ & 28257  & 0.3816538870 & 0.3940433946 & 0.3949032823 \\
8  & $10^{-15}$ & 88248  & 0.3816541694 & 0.3940419645 & 0.3949026788 \\
8  & $10^{-18}$ & 222570 & 0.3816541684 & 0.3940419569 & 0.3949026664 \\
8  & $10^{-21}$ & 468147 & 0.3816541685 & 0.3940419568 & 0.3949026663 \\
16 & $10^{-12}$ & 329045   & 0.191318    & 0.192175    & 0.191420     \\
16 & $10^{-13}$ & 600202   & 0.191188    & 0.192111    & 0.191296     \\
16 & $10^{-14}$ & 1049297  & 0.191168    & 0.192065    & 0.191274     \\
16 & $10^{-15}$ & 1773110  & 0.191154    & 0.192045    & 0.191257     \\
16 & $10^{-16}$ & 2897511  & 0.191146    & 0.192038    & 0.191250     \\
16 & $10^{-17}$ & 4581729  & 0.191143    & 0.192036    & 0.191248     \\
32 & $10^{-10}$ & 314759   & 0.097743    & 0.098547    & 0.097844     \\
32 & $10^{-11}$ & 694095   & 0.096200    & 0.096463    & 0.096015     \\
32 & $10^{-12}$ & 1443308  & 0.095490    & 0.095825    & 0.095415     \\
32 & $10^{-13}$ & 2827205  & 0.095215    & 0.095599    & 0.095159     \\
32 & $10^{-14}$ & 5228452  & 0.095122    & 0.095512    & 0.095067     \\
32 & $10^{-15}$ & 9149966  & 0.095104    & 0.095462    & 0.095050     \\
32 & $10^{-16}$ & 15161238 & 0.095098    & 0.095455    & 0.095042     \\
\end{tabular}
\label{XYSP81632}
\end{table}

\begin{table}[htbp]
\caption{
Convergence with lattice size $N$ of the crossover coupling
$g_{\text{c}}(N)$ for the $XY$ spin-Peierls model in the $\omega = 10$ case.
}
\begin{tabular}{ld}
$N$ & $g_{\text{c}}(N)$ \\
\hline
4   &  2.5308 \\
8   &  2.4338 \\
16  &  2.4049 \\
32  &  2.4083 \\
\end{tabular}
\label{gcofN}
\end{table}

\vfill
\eject

\begin{figure}[p]
\centerline{\epsfxsize=8.4cm \epsfbox{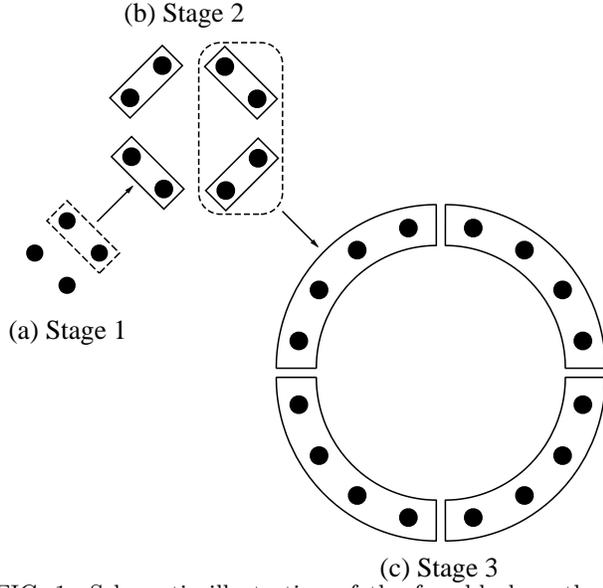}}
\caption{
Schematic illustration of the four-block method. (a) The first step is to diagonalise a four-site ring. A reduced density matrix is defined for a two-site subsystem. (b) The density matrix eigenstates are then used as a truncated basis for the two-site blocks used in the second iteration, which consists of four t-site blocks. (c) Again, a reduced density matrix is defined for a two-block (four-site) subsystem and the density matrix eigenstates are used to form a truncated basis for the four-site blocks used in the third iteration. The size of the lattice (superblock) doubles at each iteration.
}
\label{schematic}
\end{figure}

\begin{references}

\bibitem[*]{email}
Email address: ph1rb@newt.phys.unsw.edu.au

\bibitem{white}
S. R. White, Phys.\ Rev.\ Lett.\ {\bf 69}, 2863 (1992); 
Phys.\ Rev.\ B {\bf 48}, 10345 (1993).

\bibitem{review}
G. A. Gehring, R. J. Bursill and T. Xiang, Acta Physica Polonica {\bf 91}, 105 (1997);
S. R. White, Phys.\ Rep.\ {\bf 301}, 187 (1998).

\bibitem{spin1}
S. R. White and D. A. Huse, Phys.\ Rev.\ B {\bf 48}, 3844 (1993);
E. S. S{\o}renson and I. Affleck, Phys.\ Rev.\ Lett.\ {\bf 71}, 1633 (1993).

\bibitem{kondo}
C. C. Yu and S. R. White, Phys.\ Rev.\ Lett.\ {\bf 71}, 3866 (1993);
M. Guerrero and C. C. Yu, Phys.\ Rev.\ B {\bf 51}, 10301 (1995);
S. Moukouri and L. G. Caron, Phys.\ Rev.\ B {\bf 52}, 15723 (1995).

\bibitem{fermion_ladders}
R. M. Noack, S. R. White and D. J. Scalapino, Phys.\ Rev.\ Lett.\ {\bf 73}, 882 (1994);
R. M. Noack, S. R. White and D. J. Scalapino, Europhys.\ Lett.\ {\bf 30}, 163 (1995).

\bibitem{spin_ladders}
M. Azzouz, L. Chen and S. Moukouri, Phys.\ Rev.\ B {\bf 50}, 6223 (1994);
S. R. White, R. M. Noack and D. J. Scalapino, Phys.\ Rev.\ Lett.\ {\bf 73}, 886 (1994).

\bibitem{bursill1}
R. J. Bursill, T. Xiang and G. A. Gehring,  J. Phys.\ A {\bf 28}, 2109 (1994);
R. J. Bursill, G. A. Gehring, D. J. J. Farnell, J. B. Parkinson, Tao Xiang and Chen Zeng, J. Phys.\ C {\bf 7}, 8605 (1995);
R. Chitra, S. Pati, H. R. Krishnamurthy, D. Sen and S. Ramasesha,
Phys.\ Rev.\ B {\bf 52}, 6581 (1995);
K. A. Hallberg, P. Horsch and G. Martinez, Phys.\ Rev.\ B {\bf 52}, R719 (1995);
U. Schollwock and T. Jolicoeur, Europhys.\ Lett.\ {\bf 30}, 493 (1995).

\bibitem{disorder}
P. Schmitteckert and U. Eckern, Phys.\ Rev.\ B {\bf 53}, 15397 (1996);
P. Schmitteckert, T. Schulze, C. Schuster, P. Schwab and U. Eckern, Phys.\ Rev.\ Lett.\ {\bf 80}, 560 (1998).

\bibitem{dynamics}
K.\ A.\ Hallberg, Phys.\ Rev.\ B {\bf 52}, R9827 (1995);
G. Z. Wen and W. P. Su, Synth.\ Met.\ {\bf 78}, 195 (1996).

\bibitem{nishino}
T. Nishino, J. Phys.\ Soc.\ Jap.\ {\bf 64}, 3598 (1995);
T. Nishino, K. Okunishi and M. Kikuchi, Phys.\ Lett.\ A {\bf 213},
69 (1996).

\bibitem{bursill2}
R. J. Bursill, T. Xiang and G. A. Gehring, J. Phys.\ C {\bf 8}, L583 (1996);
S. Moukouri and L. G. Caron, Phys.\ Rev.\ Lett.\ {\bf 77}, 4640 (1996);
X. Wang and T. Xiang, Phys.\ Rev.\ B {\bf 56}, 5061 (1997);
D. Coombes, T. Xiang and G. A. Gehring, J. Phys.\ C {\bf 10}, L159 (1998);
T. Xiang, Phys.\ Rev.\ B {\bf 58}, 9142 (1998).

\bibitem{marston}
J. Kondev and J. B. Marston, Nuc.\ Phys.\ B {\bf 497}, 639 (1997).

\bibitem{long_range}
G. Fano, F. Ortolani and L. Ziosi, J. Chem.\ Phys.\ {\bf 108}, 9246 (1998);
D. Yaron, E. E. Moore, Z. Shuai, J. J. Bredas, J. Chem.\ Phys.\ {\bf 108}, 7451 (1998);
R. J. Bursill and W. Barford, Preprint.

\bibitem{2D_fermions}
S. Liang and H. Pang, Europhys.\ Lett.\ {\bf 32}, 173 (1995).

\bibitem{CAVO}
S. R. White, Phys.\ Rev.\ Lett.\ {\bf 77}, 3633 (1996).

\bibitem{msdmrg}
T. Xiang, Phys.\ Rev.\ B {\bf 53}, 10445 (1996).

\bibitem{white_2D}
S. R. White and D. J. Scalapino, Phys.\ Rev.\ B {\bf 55}, 14701 (1997);
Phys.\ Rev.\ B {\bf 55}, 6504 (1997);
Phys.\ Rev.\ B {\bf 57}, 3031 (1998);
Phys.\ Rev.\ Lett.\ {\bf 80}, 1272 (1998).

\bibitem{jongh}
M. S. L. du Croo de Jongh and J. M. J. van Leeuwen, Phys.\ Rev.\ B {\bf 57}, 8494 (1998).

\bibitem{compare}
J. Bonca, J. E. Gubernatis, M. Guerrero, E. Jeckelmann and S. R. White, Preprint cond-mat/9712018.

\bibitem{wetting}
E. Carlon, A. Drzewinski and J. Rogiers, Phys.\ Rev.\ B {\bf 58}, 5070 (1998).

\bibitem{quantum_chemistry}
S. R. White and R. L. Martin, Preprint cond-mat/9808118; S. R. White, Preprint cond-mat/9808293.

\bibitem{bose_Hubbard}
R. V. Pai, R. Pandit, H. R. Krishnamurthy and S. Ramasesha, Phys.\ Rev.\ Lett.\ {\bf 76}, 2937 (1996).

\bibitem{caronxy}
L. G. Caron and S. Moukouri, Phys.\ Rev.\ Lett.\ {\bf 76}, 4050 (1996).

\bibitem{acoustic_phonons}
L. G. Caron and S. Moukouri, Phys.\ Rev.\ B {\bf 56}, R8471 (1997).

\bibitem{jeckelmann}
E. Jeckelmann and S. R. White, Phys.\ Rev.\ B {\bf 57}, 6376 (1998).

\bibitem{zhang}
C. L. Zhang, E. Jeckelmann and S. R. White, Phys.\ Rev.\ Lett.\ {\bf 80}, 2661 (1998).

\bibitem{wilson}
K. G. Wilson, Rev.\ Mod.\ Phys.\ {\bf 47}, 773 (1975);
H. R. Krishna-Murthy, K. G. Wilson and J. W. Wilkins, Phys.\ Rev.\ Lett.\ {\bf 35}, 1101 (1975);
S.-T.\ Chui and J. W. Bray, Phys.\ Rev.\ B {\bf 18}, 2426 (1978);
P. Pfeuty, R. Jullien and K. A. Pearson, in {\em Real Space Renormalisation}, Topics in Current Physics, Vol.\ 30 (Springer-Verlag, Berlin, 1982);
C. Y. Pan and X. Chen, Phys.\ Rev.\ B {\bf 36}, 8600 (1987);
M. D. Kovarik, Phys.\ Rev.\ B {\bf 41}, 6889 (1990);
J. P\'{e}rez-Conde and P. Pfeuty, Phys.\ Rev.\ B {\bf 47}, 856 (1993).

\bibitem{holstein}
R. J. Bursill, R. H. McKenzie and C. J. Hamer, Phys.\ Rev.\ Lett.\ {\bf 80}, 5607 (1998).

\bibitem{lepetit}
M.-B.\ Lepetit and G. M. Pastor, Phys.\ Rev.\ B, In press.

\bibitem{bethe}
H. Bethe, Z. Phys.\ {\bf 71}, 205 (1931);
J. des Cloizeaux and M. Gaudin, J. Math.\ Phys.\ {\bf 7}, 1384 (1966);
D. C. Mattis (ed.), {\em The Many Body Problem: An
Encyclopedia of Exactly Solvable Models in One Dimension}, (World Scientific, Singapore, 1993).

\bibitem{finite_rings}
L. V. Andreev, J. Phys.\ A {\bf 23}, L485 (1990).

\bibitem{bulaevskii}
L. N. Bulaevskii, Sov.\ Phys.\ JETP {\bf 17}, 1008 (1963).

\bibitem{baxter}
R. J. Baxter, J. Phys.\ C {\bf 6}, L94 (1973).

\bibitem{footnote}
For an isotropic system $\Delta_{\text{dimer}}$ is the singlet-singlet gap and the two doublet states $\Delta_{\text{doublet}}$ are degenerate with $\Delta_{\text{N\'{e}el}}$---the three excitations forming the lowest-lying triplet.

\bibitem{nomura1}
K. Okamoto and K. Nomura, Phys.\ Lett.\ A {\bf 169}, 433 (1992).

\bibitem{nomura2}
K. Nomura and K. Okamoto, J. Phys.\ A {\bf 27}, 5773 (1994).

\bibitem{bursillX}
R. J. Bursill, R. H. McKenzie and C. J. Hamer, Unpublished.

\end{references}
\end{document}